\documentclass[12pt]{iopart}
\usepackage{graphicx}
\usepackage[caption=false]{subfig}
\usepackage{epstopdf}
\usepackage{amsfonts}

\begin{document}


   \title{Back to epicycles -- relativistic Coulomb systems in velocity space}

   \author{Uri Ben-Ya'acov}

   \address{School of Engineering, Kinneret Academic College on
   the Sea of Galilee, \\   D.N. Emek Ha'Yarden 15132, Israel}

   \ead{uriby@kinneret.ac.il}


\begin{abstract}
The study of relativistic Coulomb systems in velocity space is prompted by the fact that the study of Newtonian Kepler/Coulomb systems in velocity space, although less familiar than the analytic solutions in ordinary space, provides a much simpler (also more elegant) method. The simplicity and elegance of the velocity-space method derives from the linearity of the velocity equation, which is the unique feature of $1/r$ interactions for Newtonian and relativistic systems alike. The various types of possible trajectories are presented, their properties deduced from the orbits in velocity space, accompanied with illustrations. In particular, it is found that the orbits traversed in the relativistic velocity space (which is hyperbolic ($H^3$) rather than Euclidean) are epicyclic -- circles whose centres also rotate -- thus the title. \footnote{Dedicated to the memory of J. D. Bekenstein -- physicist, teacher and human}
\end{abstract}



\vskip30pt

\section{Introduction}

The motivation for the present work stems from the success of studying Newtonian Kepler/Coulomb (KC) 2-body systems in velocity space. This method was originally presented by Hamilton in 1847 \cite{Hamilton1847}, and elaborated years later mainly by Maxwell \cite{Maxwell} and Feynmann \cite{Goodstein96}. It discusses the dynamics of the system by following the orbit traced by the tip of the velocity vector (termed {\it hodograph} by Hamilton). Although hardly known, the virtue of the hodograph method is that its application to classical systems with $1/r$ potential provides, in a very simple and elegant way, the full solution -- all the necessary information regarding the dynamics of the system, including spatial trajectories -- just from the discussion in velocity space. Its merits have been discusses on several occasions in the last decades \cite{Milnor1983,GonVilla.etal,Butikov2000,Derbes2001,KowenMathur2003,Munoz2003,Carinena.etal2016}.

The success of the hodograph method with the Newtonian KC systems triggers an interest in its possible application to relativistic systems. The simplest extension of Newtonian KC systems to relativity are relativistic Coulomb systems -- the limit of EM 2-body systems when one of the charges is much heavier than the other. Such systems were studied in the literature (\cite{Boyer2004,Stahlhofen2005a} and references therein), generally following standard methods in configuration space. The present work complements these studies, with a thorough discussion in relativistic velocity space.

The essentials of the hodograph method are presented in the following. It is worthwhile to emphasize that the hodograph method, Newtonian and relativistic alike, is not a study in Hamiltonian phase space $\left\{\left(x,p\right)\right\}$, but an analysis confined only to the velocities, making use of the rotational symmetry of the system and the associated angular-momentum conservation to transform spatial dependencies to velocity dependencies. The use of rotational symmetry in this way is certainly an essential part of the method. Also it is to be noted that the relativistic velocity space is a 3-D hyperboloid $H^3$ embedded in a 1+3 pseudo-Euclidean space \cite{RhodesSemon2004} (unlike the 3-D Euclidean velocity space for Newtonian systems), and the results depend heavily on this property.

The talk started with reviewing the classical KC hodograph method (see section \ref{sec: hodKC}), followed with an account of the relativistic velocity space, the relativistic hodograph equations and general properties of the hodographs in section \ref{sec: hodrelC}. The main body of the talk then consists of an illustrative account of the various particular cases.

A fuller account of the results is in preparation to be published elsewhere.

{\it Notation}. The convention $c=1$ is used throughout, unless specified otherwise. Events in Minkowski space-time are $x^\mu = \left(x^0,x^1,x^2,x^3\right)$, with metric tensor $g_{\mu\nu} = {\rm diag} \left(-1,1,1,1\right) \, , \, \mu,\nu = 0,1,2,3$. For any 4-vectors $a^\mu = (a^0,\vec a)$ and $b^\mu = (b^0,\vec b)$, their inner product is then $a \cdot b = -a^0 b^0 + \vec a \cdot \vec b$.

\vskip20pt

\section{The hodograph in classical Kepler/Coulomb systems} \label{sec: hodKC}

Classical KC systems are determined by the Hamiltonian
 \begin{equation} \label{eq: classHam}
 H\left(\vec r, \vec p \right) = \frac{{\vec p \,}^2}{2m}  + \frac{\kappa}{r}
 \end{equation}
with the velocity equation of motion and the conserved angular momentum
 \begin{equation} \label{eq: peqmot}
 m\frac{d\vec v}{dt} = \frac{\kappa}{r^3} \vec r \quad , \quad  \ell = m r^2 \frac{d\theta}{dt} \, .
 \end{equation}
Coordinates $x,y$ are assumed the for the plane of motion, with the conserved angular momentum vector $\vec \ell  = \vec r \times \vec p = \ell \hat z$ perpendicular to it. Polar coordinates $(r,\theta)$ are also used, with the polar-planar unit vectors
 \begin{equation} \label{eq: poluvec}
 \hat r = \cos \theta \hat x + \sin \theta \hat y  \quad ,  \quad  \hat \theta  =  - \sin \theta \hat x + \cos \theta \hat y
 \end{equation}
satisfying
 \begin{equation} \label{eq: poluvec2}
 \hat r = - \hat z \times \hat \theta = - \frac{d \hat\theta}{d\theta}  \quad , \quad  \hat \theta = \hat z \times \hat r = \frac{d \hat r}{d\theta} \, .
 \end{equation}
Eliminating the time variable from both equations in \eref{eq: peqmot} yields the velocity-angular equation
\begin{equation}\label{eq: veqangKC}
 \frac{d\vec v}{d\theta} = \frac{\kappa}{\ell} \hat r  =  - \frac{\kappa}{\ell}\frac{d\hat \theta}{d\theta}  \, ,
\end{equation}
with the general hodograph solution
 \begin{equation} \label{eq: hodoN}
\vec v \left(\theta\right) = \vec B_o - \frac{\kappa}{\ell} \hat \theta \, .
 \end{equation}
The constant of integration $\vec B_o$ is known as {\it Hamilton's vector}. The solution \eref{eq: hodoN} describes a circle (or at least a circular arc for unbound systems) in velocity space, centred around $\vec B_o$ with radius $|\kappa| / \ell$.

\begin{figure}
  \centering
  \includegraphics[width=10cm]{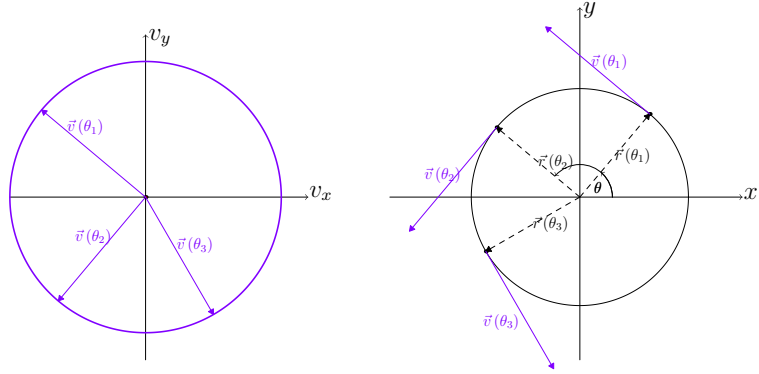}
  \caption{\label{fig:1}Newtonian hodograph method : For spatial circular motion (right) the hodograph (left) is simply a circle centred at the origin of the velocity space.}
\end{figure}
\begin{figure}
  \centering
  \includegraphics[width=10cm]{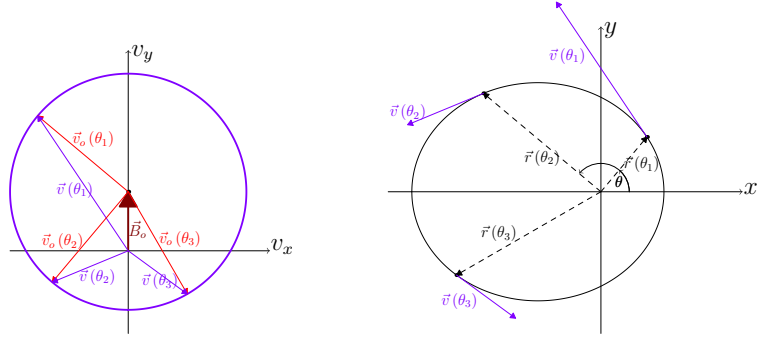}
  \caption{\label{fig:2}Newtonian hodograph method : For spatial bound non-circular motion (elliptic -- right) the hodograph (left) is still a circle but shifted from the origin of the velocity space.}
\end{figure}

The spatial trajectory $\vec r \left(\theta\right) = r\left(\theta\right) \hat r\left(\theta\right)$ derives from \eref{eq: hodoN} using the relation
 \begin{equation} \label{eq: vtheta}
 v_\theta = \vec v \cdot \hat \theta = \frac{\ell }{mr} \, ,
 \end{equation}
from which also follows the energy equation, completely confined to velocity space variables,
\begin{equation}\label{eq: energyintN}
 \frac{m {\vec v \,}^2}{2} + \frac{\kappa}{r} = \frac{m {\vec v \,}^2}{2} + \frac{m \kappa}{\ell} v_\theta = E\,' \, .
\end{equation}

\begin{figure}
  \centering
  \subfloat[Attraction. \label{fig:3a}] {\includegraphics[width=10cm]{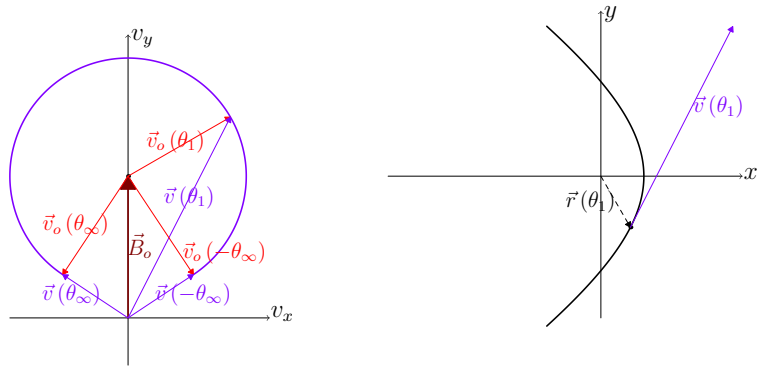}} \\
  \subfloat[Repulsion. \label{fig:3b}] {\includegraphics[width=10cm]{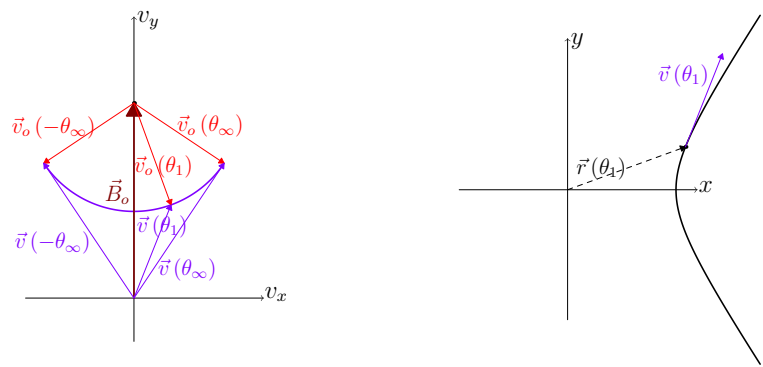}}
  \caption{\label{fig:3} Newtonian hodograph method : For spatially unbound motion (right) the hodographs (left) are finite circular arcs with the centre shifted from the origin of the velocity space. The velocity vectors are tangent to the arcs at both ends.}
\end{figure}

The nature of the solution depends on Hamilton's vector. Substituting \eref{eq: hodoN} in \eref{eq: energyintN}, the relation
\begin{equation}\label{eq: BoNewt}
 {B_o}^2 = \frac{2E\,'}{m} + \frac{\kappa^2}{\ell^2} \, ,
\end{equation}
follows. The minimal energy hodograph, with $\vec B_o = 0$, is a canonical circle $C_o(\ell)$ drawn by the tips of the vectors $\vec v_o \left(\theta\right) = -\left(\kappa/\ell\right) \hat \theta$ centred at the origin of velocity space (\Fref{fig:1}), corresponding to circular spatial motion. With increasing energy $\vec B_o$ becomes non-zero, the corresponding hodographs \eref{eq: hodoN} remain circles with the same radius $\left |\kappa\right| / \ell$ but shifted from the origin of velocity space by $\vec B_o$ (\Fref{fig:2} \& \ref{fig:3}), and the spatial orbits become conic sections. In particular, for unbound systems the hodographs reduce the infinite spatial trajectory into a finite circular arc (\Fref{fig:3}). The endpoints correspond to $v_\theta = 0 \Leftrightarrow r \rightarrow \infty$, where $\vec v$ is tangent to the hodograph.

The interested reader may find more details regarding the hodograph method in classical systems in several publications \cite{Milnor1983,GonVilla.etal,Butikov2000,Derbes2001,KowenMathur2003,Munoz2003,Carinena.etal2016}.

\vskip20pt

\section{The hodograph method in relativistic Coulomb systems} \label{sec: hodrelC}

We use for the variables of the relativistic velocity space unit velocity (time-like future-directed) 4-vectors $u^\mu = \gamma(v) \left(1 , \vec v \right)$, all satisfying $u \cdot u = - 1$. The relativistic velocity space is then the space of all unit velocity 4-vectors, namely the 3-D unit hyperboloid ($H^3$) \cite{RhodesSemon2004}
\begin{equation}\label{eq: Vrel}
 {\cal V}_{\rm rel} \equiv \left\{ u^\mu = \left(u^0 , \vec u \right) | u^0 = \sqrt{1 + \vec u\, ^2} \right\}
\end{equation}
embedded in a 4-D pseudo-Euclidian space
\begin{equation}\label{eq: E13}
E^{(1,3)} = \left\{ w^\mu = \left( w^0,\vec w \right) \in \mathbb{R}^4 | g_{\mu\nu} = {\rm diag} \left(-1,1,1,1\right) \right\} \, .
\end{equation}

Relativistic Coulomb systems are determined by the Hamiltonian
 \begin{equation} \label{eq: Ham}
 H\left(\vec r, \vec p \right) = \sqrt {{\vec p \,}^2 + m^2}  + \frac{\kappa}{r}
 \end{equation}
The hodograph equations that follow from Hamilton's equations may be formulated for all $w^\mu \in E^{(1,3)}$ :
 \begin{equation} \label{eq: weqns}
 \frac{d\vec w}{d\theta} = \frac{\kappa}{\ell} w^0 \hat r \quad , \quad \frac{dw^0}{d\theta} = \frac{\kappa}{\ell} \vec w \cdot \hat r
 \end{equation}
and may be combined Lorentz-covariantly as
 \begin{equation} \label{eq: weqns2}
 \frac{dw^\mu}{d\theta} = \frac{\kappa}{\ell}{\Omega^\mu}_\nu w^\nu
 \end{equation}
with the 4-D rotation matrix
\begin{equation}\label{eq: Omega}
{\Omega^\mu}_\nu = \left(\begin{array}{*{20}{c}}
  0 & \vline & {\hat r} \\
\hline
  {\hat r} & \vline & 0
\end{array}
\right) = \left(\begin{array}{*{20}{c}}
  0 & \vline & {\cos \theta} & {\sin \theta} & 0 \\
\hline
  {\cos \theta} & \vline & 0 & 0 & 0 \\
  {\sin \theta} & \vline & 0 & 0 & 0 \\
  0 & \vline & 0 & 0 & 0
\end{array}
\right)
\end{equation}
Since $\Omega_{\mu\nu} = -\Omega_{\nu\mu}$, it follows that for any two solutions $w_1^\mu \left( \theta \right),w_2^\mu \left( \theta \right)$ of \eref{eq: weqns2} in $E^{(1,3)}$ the inner product $w_1 \cdot w_2$ is constant. In particular, any solution of the hodograph equations in $E^{(1,3)}$ is of constant magnitude.

The hodograph equations induce pseudo-rotation in $E^{(1,3)}$, relative to an axis that itself rotates with $\theta$. The axis of rotation is identified as being along the vector
\begin{equation}\label{eq: vodef}
 v_o^\mu = \left( 1, - \frac{\kappa}{\ell} \hat \theta \right)
\end{equation}
It is easily verified that the vector \eref{eq: vodef} is preserved by \eref{eq: weqns2}. The fact that the classical basic circle $\vec v_o = - \left(\kappa/\ell\right) \hat\theta$ is the spatial component of $v_o^\mu$ is very significant and will be discussed in the following.

The hodographs of relativistic Coulomb systems are the solutions of \eref{eq: weqns2} that are confined to ${\cal V}_{\rm rel}$. The confinement is insured by the constant norm of the solution. Such general solutions, depending on the total energy $E$ and angular momentum $\ell$ as parameters, are denoted in the following $u^\mu \left( \theta |E,\ell \right)$. The principal significance of the axis vector \eref{eq: vodef} is then in determining the energy integral, and thus the energy dependence of the hodograph :
\begin{equation}\label{eq: enerint}
 - u \cdot {v_o} = {u^0} + \frac{\kappa }{\ell }{u_\theta } = \frac{E}{m}
\end{equation}
($u_\theta = \vec u \cdot \hat \theta$), so the energy is the (constant) projection of the kinematical momentum $m u^\mu$ on the rotation axis.

It is fairly straight-forward to obtain the general energy dependence of the hodographs. Consider the infinitesimal variation between two hodographs whose energy differ by $\delta E$,
\begin{equation}\label{eq: hodovar}
 \delta u^\mu = u^\mu \left( \theta | E + \delta E, \ell\right) - u^\mu \left( \theta | E, \ell\right) = \frac{\partial u^\mu}{\partial E} \left( \theta |E,\ell \right)\delta E
\end{equation}
The constraints $u \cdot v_o = - E/m$ and $u \cdot u = -1$ imply that
\begin{equation}\label{eq: deludelE}
 \frac{\partial u^\mu}{\partial E} = \frac{E u^\mu - m v_o^\mu}{\Lambda^2 \left( E,\ell \right)}
\end{equation}
with
\begin{equation}\label{eq: defLam}
 \Lambda^2 \left( E,\ell \right) = \left(E u - m v_o\right)^2 = E^2 + m^2 \left(\frac{\kappa^2}{\ell^2} - 1\right) \, .
\end{equation}
The vector $\partial u^\mu / \partial E$ is a space-like 4-vector, being orthogonal to $u^\mu$ and tangent to $\mathcal{V}_{\rm rel}$ at $u^\mu$, therefore $\Lambda^2 \left( E,\ell \right) = \left(\partial u / \partial E\right)^{-2} > 0$. Integrating \eref{eq: deludelE} as a 1st order ODE in $E$ then yields the generic form of the hodographs,
\begin{equation}\label{eq: genhod}
 u^\mu \left( \theta |E,\ell \right) = w_o^\mu \left( \theta |E,\ell \right) + \Lambda\left(E,\ell\right) n_o^\mu \left( \theta |\ell \right)
\end{equation}
where
\begin{equation}\label{eq: wforgenhod}
 w_o^\mu \left( \theta |E,\ell \right) = \cases{ \frac{E}{\left( 1 - {\raise0.7ex\hbox{${\kappa^2}$} \!\mathord{\left/
 {\vphantom {\kappa^2 \ell^2}} \right.\kern-\nulldelimiterspace} \!\lower0.7ex\hbox{$\ell^2$}} \right) m} v_o^\mu \left( \theta |\ell \right) & for $\ell \ne \left| \kappa \right|$\\
 \frac{m}{2E} v_o^\mu \left( \theta |\ell \right) & for $\ell = \left| \kappa \right|$\\} \, ,
\end{equation}
and $n_o^\mu \left( \theta |\ell \right)$ is an energy-independent solution of the hodograph equation \eref{eq: weqns2} (whose explicit form, depending on $\kappa/\ell$, will be given in the following), and also linearly independent of $v_o^\mu$. Equation \eref{eq: genhod} demonstrates $u^\mu \left( \theta |E,\ell \right)$ as the linear combination of an on-axis component $w_o^\mu$ and an off-axis component
\begin{equation}\label{eq: genB}
 B^\mu \left( \theta |E,\ell \right) \equiv \Lambda\left(E,\ell\right) n_o^\mu \left( \theta |\ell \right) \, .
\end{equation}
Both have constant magnitudes but changing directions. Being both solutions of the hodograph equation \eref{eq: weqns2}, they also maintain constant relative angle. The combined motion is therefore epicyclic, or precession -- a rotation ($B^\mu$) imposed on another rotation ($w_o^\mu$). $B^\mu \left( \theta |E,\ell \right)$ will be recognized in the following as the {\it relativistic Hamilton vector}.

Once the hodograph solution $u^\mu \left( \theta |E,\ell \right)$ is available, it is also straight-forward to obtain the corresponding spatial trajectory. In simile to \eref{eq: vtheta}, from the definition of the conserved angular momentum $\ell = mr u_\theta$ then follows
 \begin{equation} \label{eq: rtheta}
 \vec r (\theta) = \frac{\ell}{m u_\theta(\theta)} \hat r(\theta) \, .
 \end{equation}
In return, since $r > 0$, the requirement that $u_\theta$ must be non-negative, $u_\theta(\theta) \ge 0$, with equality possible only for unbound configurations at $ r \to \infty$, determines the possible $\theta$-range.

\begin{figure}[h]
  \centering
  \includegraphics[width=6cm]{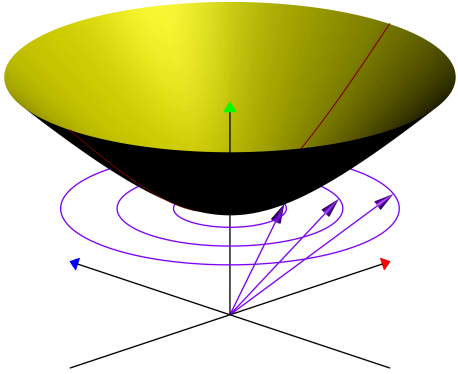}\hspace{20pt}%
  \begin{minipage}[b]{250pt}
  \caption{\label{fig:5} Relativistic hodograph method : The axis vector $v_o^\mu$ and the basis circle $C_o(\ell)$ that it traverses, relative to the velocity space ${\cal V}_{\rm rel}$ hyperboloid, for the 3 possible cases :
  (i) left arrow -- $v_o^\mu$ time-like (for $\ell > |\kappa|$), if continued the vector punches the hyperboloid into its interior;
  (ii) middle arrow -- $v_o^\mu$ light-like (for $\ell = |\kappa|$), if continued the vector approaches the hyperboloid asymptotically;
  (iii) right arrow -- $v_o^\mu$ space-like (for $\ell < |\kappa|$), if continued the vector recedes from the hyperboloid.}
  \end{minipage}
\end{figure}

The ratio $|\kappa| / \ell$ determines the nature of the rotation axis, and consequently of the hodograph itself. From the magnitude of the axis vector \eref{eq: vodef} we get
\begin{equation}\label{eq: vomag}
{v_o}^2 = \frac{\kappa^2}{\ell^2} - 1 \quad \begin{array}{*{20}{c}}
 \nearrow \\
 \to \\
 \searrow
\end{array} \quad \begin{array}{*{20}{c}}
{\ell > |\kappa|}&{\Leftrightarrow}&{v_o^\mu \, {\textrm {time-like}}}\\
{\ell = |\kappa|}&{\Leftrightarrow}&{v_o^\mu \, {\textrm {light-like}}}\\
{\ell < |\kappa|}&{\Leftrightarrow}&{v_o^\mu \, {\textrm {space-like}}}
\end{array}
\end{equation}
$|\kappa| / \ell$ is the magnitude of the spatial velocity on the basis circle -- minimal energy circle -- formed by $v_o^\mu$, $C_o(\ell) = \left\{v_o^\mu(\theta)\right\}$ (see \Fref{fig:5}). The relativistic requirement that particles' velocities cannot reach the velocity of light necessarily implies that the relativistic solution can have non-relativistic limit only for time-like $v_o^\mu$. The following results split accordingly.

A note regarding the hodograph illustrations in the following : Since the relativistic hodographs are 3-D curves, each of the cases discussed is illustrated with 3 projected views of a segment of the hodograph -- from above, from the side (horizontal) and an oblique-diagonal view -- accompanied by an illustration of the corresponding spatial orbit. All these illustrations share a common colour code : The axes with red, blue and green arrowhead correspond, respectively, to $u_x$, $u_y$ and $u^0$ (the $u_z$ axis suppressed); the thick red line is the hodograph itself; the blue circle is the basis circle $C_o\left(\ell\right) = \left\{v_o^\mu\left(\theta\right)\right\}$; brown is the colour for the vector $w_o^\mu$ and the circle $\left\{w_o^\mu\left(\theta\right)\right\}$ it draws; the orange line is a cross-section of the velocity space hyperboloid ${\cal V}_{\rm{rel}}$ in the $\left(u_x = u_y\right)\textrm{-}u^0$ plane; and the black line is the Hamilton vector in each case. These colours cannot be seen in the printed version, and the reader is advised to use the on-line or PDF versions. The scales are different according to case.

\vskip20pt

\section{Hodographs with Newtonian limit ($\ell > \left|\kappa\right|$)} \label{sec: ellarge}

For $\ell > |\kappa|$ the solution of the hodograph equation \eref{eq: weqns2} (with an arbitrary shift angle put to zero) is
\begin{equation}\label{eq: usol}
\begin{eqalign}
 {\vec u &=  B_o \sin \left( \beta \theta \right) \hat r + \left[ - \frac{\kappa E}{m \ell \beta^2} + \frac{B_o}{\beta} \cos \left( \beta \theta \right) \right] \hat \theta \, , \\
 u^0 &= \frac{E}{m} - \frac{\kappa}{\ell} u_\theta = \frac{E}{m \beta^2} - \frac{\kappa B_o}{\beta\ell} \cos \left( \beta \theta \right)}
\end{eqalign}
\end{equation}
with the notations
\begin{equation}\label{eq: betaBo}
 \beta = \sqrt {1 - \frac{\kappa^2}{\ell^2}} \quad , \quad B_o = \sqrt {\frac{E^2}{m^2\beta^2} - 1} = \frac{\Lambda\left(E,\ell\right)}{m \beta} \, .
\end{equation}
It may be cast into the generic hodograph form \eref{eq: genhod} as
\begin{equation}\label{eq: usol2}
 u^\mu\left(\theta|E,\ell\right) = \frac{E}{m\beta^2} v_o^\mu + B_o n_1^\mu
 \end{equation}
with the vector
\begin{equation}\label{eq: n1def}
 n_1^\mu\left(\theta|\ell\right) \equiv \left( - \frac{\kappa}{\beta \ell} \cos \left( \beta \theta \right), \sin \left( \beta \theta \right) \hat r + \frac{1}{\beta} \cos \left( \beta \theta \right) \hat \theta \right)
\end{equation}
which is a space-like unit 4-vector, tangent to $\mathcal{V}_{\rm rel}$ at $u_o^\mu$, satisfying $n_1 \cdot n_1 = 1 \, , \, u_o \cdot n_1 = 0$.

As is evident from \Fref{fig:5}, the axis vector $v_o^\mu$, if continued, punches the velocity space hyperboloid. The punching point is at $u_o^\mu\left(\theta|\ell\right) = \beta^{-1}v_o^\mu\left(\theta|\ell\right)$, which is the hodograph that corresponds, for any given value of $\ell > \left|\kappa\right|$, to the state of minimal energy with $E_{\min} = \beta m$ and $B_o = 0$, describing a uniform circular orbit in $\mathcal{V}_{\rm rel}$ and corresponding to spatial circular motion.

For $E > \beta m$, the vector $w_o^\mu = \left(E/m\beta^2\right) v_o^\mu$, being time-like, punching through the velocity hyperboloid ${\cal V}_{\rm rel}$ at $u_o^\mu$, describes a circle which lies horizontally within ${\cal V}_{\rm rel}$, with its centre right above the origin of the embedding space $E^{(1,3)}$. Then, adding $B_o n_1^\mu$ at the tip of $w_o^\mu$ (and recalling that both are orthogonal as 4-vectors), the superposed vector reaches ${\cal V}_{\rm rel}$ at $u^\mu$.

The representation \eref{eq: usol2} strongly resembles the Newtonian solution \eref{eq: hodoN} : Not only is $B_o n_1^\mu$ superposed upon the uniformly rotating vector $w_o^\mu$, but in the nonrelativistic limit $w_o^\mu \to \left(1,\vec v_o\right)$ and $B_o n_1^\mu \to \left(0,\vec B_o\right)$, the latter thus reducing to the classical Hamilton vector. It seems therefore appropriate to regard $B^\mu \equiv B_o n_1^\mu$ as the {\it relativistic Hamilton vector}.

The vector $B^\mu$ is not constant, but rotates with constant magnitude. Its derivative tangent to ${\cal V}_{\rm rel}$ at $u_o^\mu$ follows from
\begin{equation}\label{eq: n1eq}
  \frac{Dn_1^\mu}{d\theta} = \frac{dn_1^\mu}{d\theta} - \left( n_1 \cdot \frac{d u_o}{d\theta} \right) u_o^\mu = - \frac{\kappa^2}{\beta \ell^2} n_2^\mu \, .
\end{equation}
where
\begin{equation} \label{eq: n2def}
 n_2^\mu \left(\theta|\ell\right) = \left( \frac{\kappa}{\beta \ell} \sin \left( \beta \theta \right), \cos \left( \beta \theta \right) \hat r - \frac{1}{\beta} \sin \left( \beta \theta \right) \hat \theta \right)
\end{equation}
is another unit 4-vector tangent to ${\cal V}_{\rm rel}$ at $u_o^\mu$, orthogonal to $n_1^\mu\left(\theta|\ell\right)$ and satisfying a similar equation,
\begin{equation}\label{eq: n2eq}
  \frac{Dn_2^\mu}{d\theta} = \frac{dn_2^\mu}{d\theta} - \left( n_2 \cdot \frac{d u_o}{d\theta} \right) u_o^\mu = \frac{\kappa^2}{\beta \ell^2} n_1^\mu \, .
\end{equation}
It follows that $B^\mu$ rotates uniformly around the axis of rotation.

\subsection{Bound states}

\begin{figure}
  \centering
  \subfloat[Hodograph -- side view \label{fig:6a}]{\includegraphics[width=4.5cm]{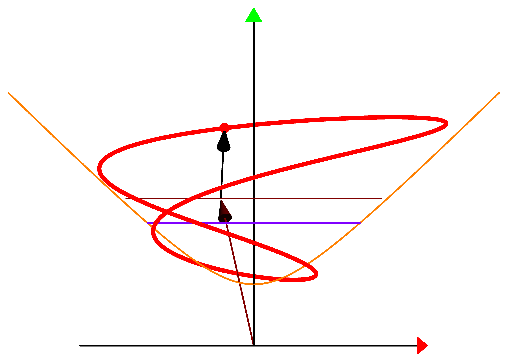}} \qquad
  \subfloat[Hodograph -- oblique view \label{fig:6b}]{\includegraphics[width=4.5cm]{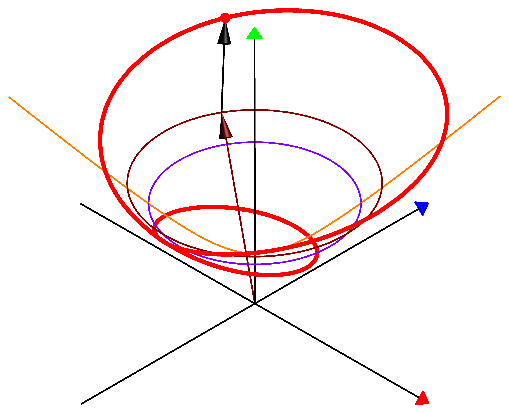}} \\
  \subfloat[Hodograph -- above view \label{fig:6c}]{\includegraphics[width=4.5cm]{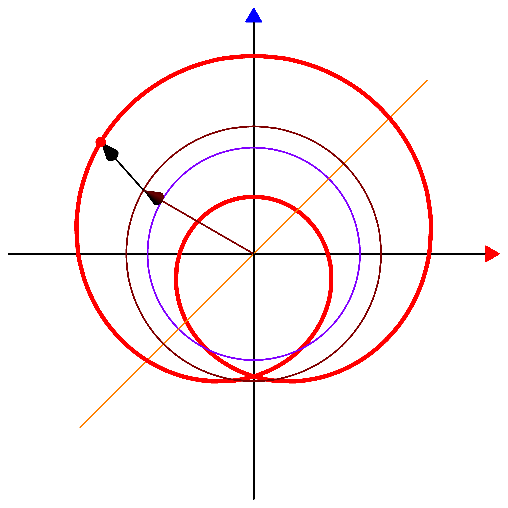}} \qquad
  \subfloat[Spatial orbit \label{fig:6d}]{\includegraphics[width=4.5cm]{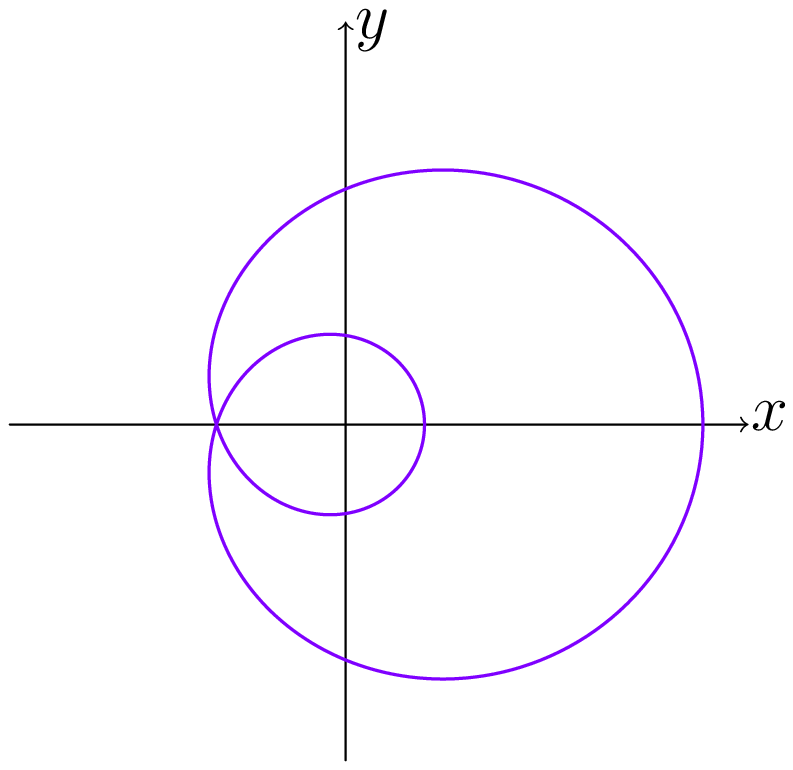}}
  \caption{\label{fig:6} Hodograph and spatial orbit for the bound state for $v_o^\mu$ time-like ($E = 0.6m, \kappa/\ell = -\sqrt{3}/2$). The trajectory is closed (periodical) because in this particular case $\beta = 1/2$.}
\end{figure}

Bound states are obtained only for $\kappa < 0$ (attraction) and $\beta m \le E < m$. Extrema are obtained when $u_r = 0$, namely $\beta \theta = n\pi \, , \, n \in \mathbb{Z}$, with
 \begin{equation} \label{eq: utboundex}
 \frac{\left| \kappa \right|E}{m\ell \beta^2} - \frac{B_o}{\beta} \le u_\theta \le \frac{\left| \kappa \right|E}{m\ell \beta^2} + \frac{B_o}{\beta}
 \end{equation}
$u_\theta$ cannot vanish, thus determining, via the relation $u_\theta = \ell/mr$, corresponding lower and upper bounds for $r$ :
 \begin{equation} \label{eq: rminmax}
 \frac{\left| \kappa \right| E - \ell \Lambda }{m^2 - E^2} \le r \le \frac{\left| \kappa \right| E + \ell \Lambda }{m^2 - E^2} \, .
 \end{equation}
Since $\beta \ne 1$ the hodograph is not periodic $2\pi$ in $\theta$, therefore not circular and the spatial trajectories are rotating ellipses. Periodicity is obtained only when $\beta$ is a rational number.

An exemplary hodograph, using the values $E = 0.6m, \, \kappa/\ell = -\sqrt{3}/2 \, (\beta = 0.5)$ together with the corresponding spatial orbit obtained via \eref{eq: rtheta} is demonstrated in \Fref{fig:6}.

\subsection{Unbound states}

\begin{figure}
  \centering
  \subfloat[Hodograph -- side view \label{fig:8a}]{\includegraphics[width=4.5cm]{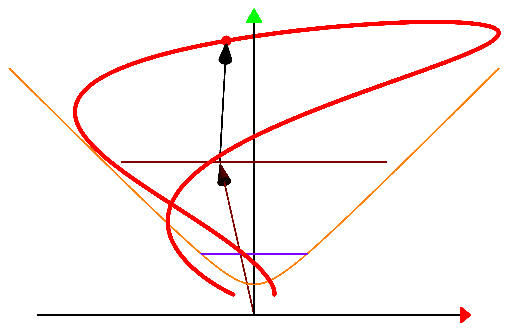}} \qquad
  \subfloat[Hodograph -- oblique view \label{fig:8b}]{\includegraphics[width=4.5cm]{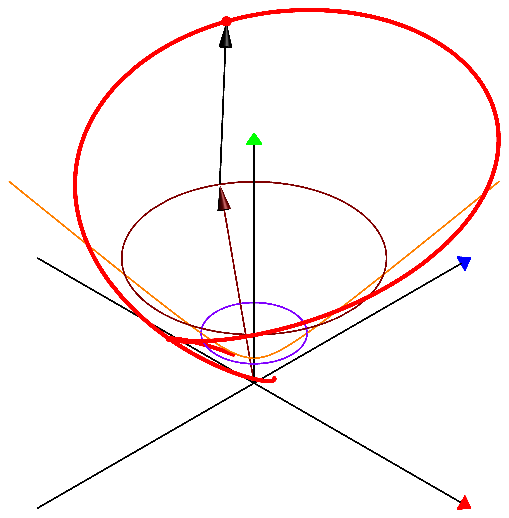}} \\
  \subfloat[Hodograph -- above view \label{fig:8c}]{\includegraphics[width=4.5cm]{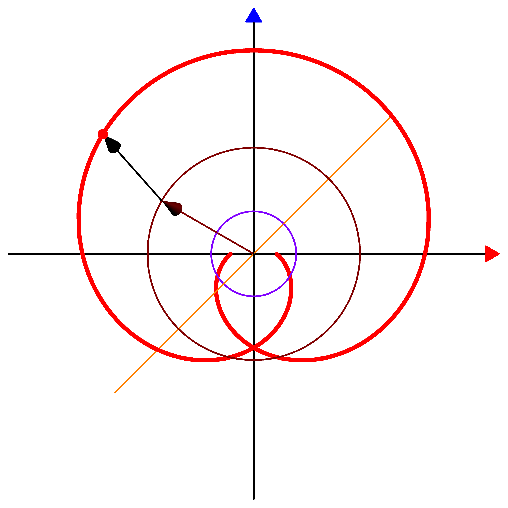}} \qquad
  \subfloat[Spatial orbit \label{fig:8d}]{\includegraphics[width=4.5cm]{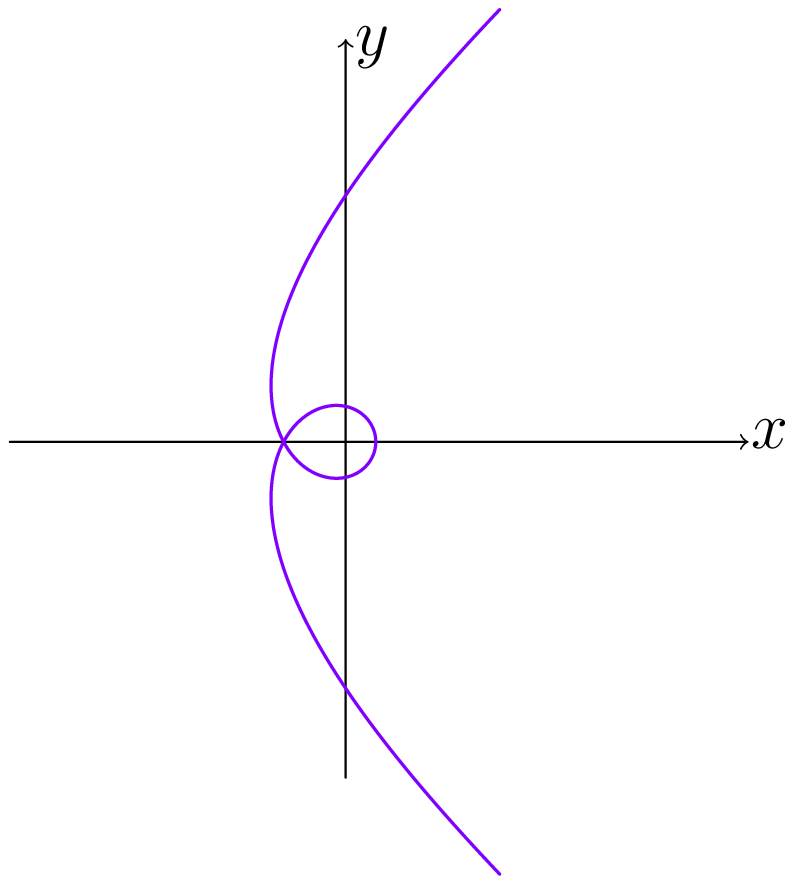}}
  \caption{\label{fig:8} Hodograph and spatial orbit for the attraction unbound state for $v_o^\mu$ time-like ($E = 1.25m, \kappa/\ell = -\sqrt{3}/2$).}
\end{figure}

For unbound systems, with $E \ge m$, the infinite limit $r \to \infty$ corresponds, according to \eref{eq: rtheta}, to $u_\theta = 0$. The infinite spatial trajectory turns into a finite-size orbit in velocity space, another merit of the hodograph method. The finite angular range of the spatial trajectory is $-\theta_\infty < \theta < \theta_\infty$, with the endpoints (at which $u_\theta = 0$) at
\begin{equation} \label{eq: thetainfty}
 \theta_\infty = \cases{
 \frac{\psi_\infty}{\beta} & $\kappa > 0$ \\
 \frac{\pi - \psi_\infty}{\beta} & $\kappa < 0$ \\}
 \quad , \quad \psi_\infty = \sin^{-1} \left( \frac{\sqrt{E^2 - m^2}}{mB_o} \right)
\end{equation}
The distance of closest approach is also found from $\max \left({u_\theta}\right)$ via \eref{eq: rtheta},
 \begin{equation} \label{eq: rmin1}
 r_{\min} = \frac{\ell \Lambda - \kappa E}{E^2 - m^2}
 \end{equation}

\begin{figure}
  \centering
  \subfloat[Hodograph -- side view\label{fig:7a}]{\includegraphics[width=4.5cm]{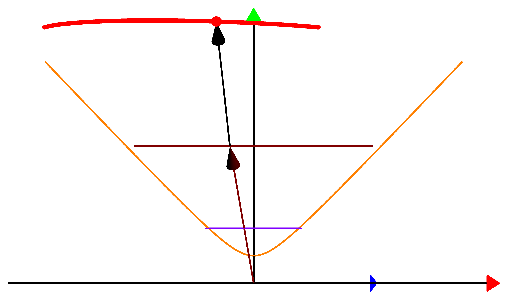}} \qquad
  \subfloat[Hodograph -- oblique view\label{fig:7b}]{\includegraphics[width=4.5cm]{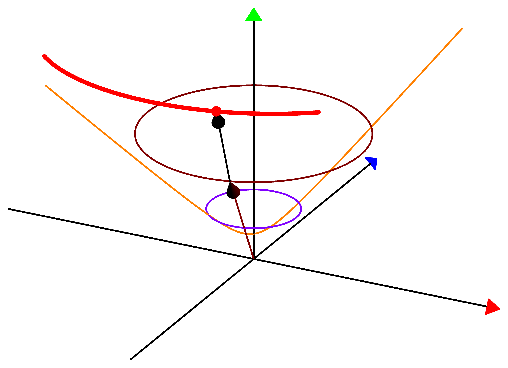}} \\
  \subfloat[Hodograph -- above view\label{fig:7c}]{\includegraphics[width=4.5cm]{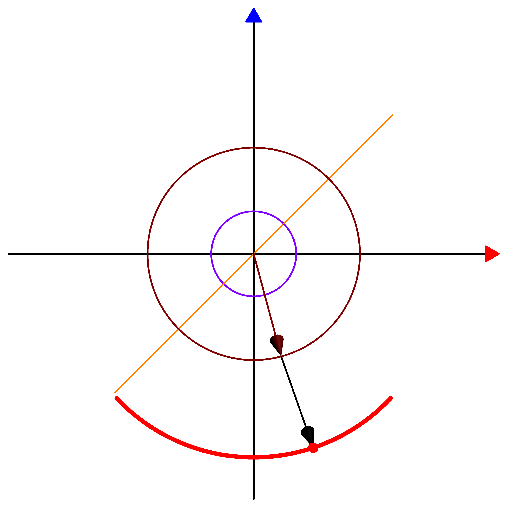}} \qquad
  \subfloat[Spatial orbit \label{fig:7d}]{\includegraphics[width=4.5cm]{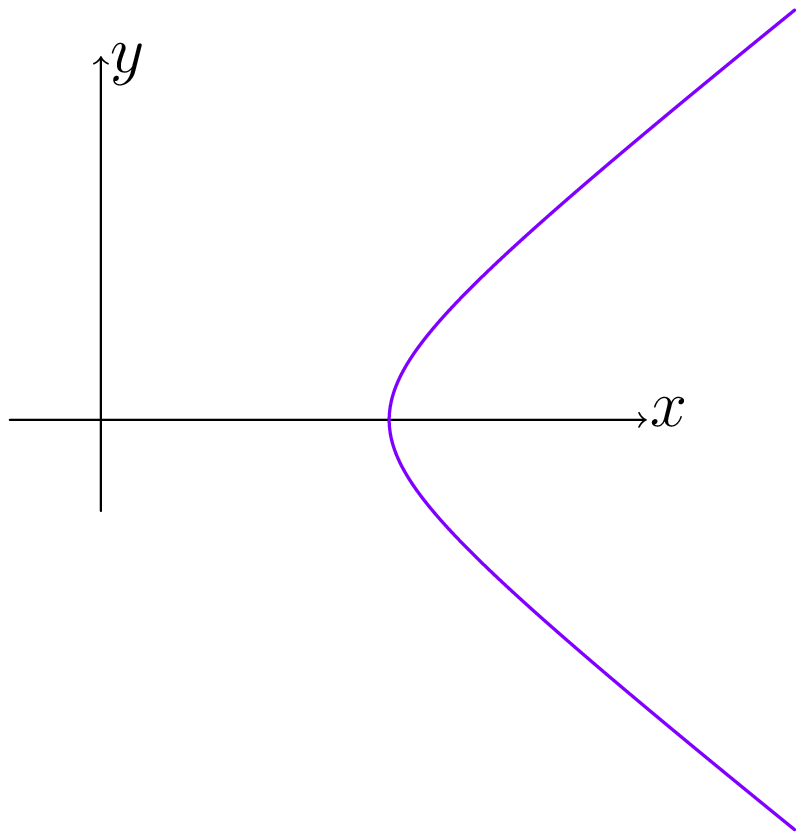}}
  \caption{\label{fig:7} Hodograph and spatial orbit for the repulsion unbound state for $v_o^\mu$ time-like ($E = 1.25m, \kappa/\ell = \sqrt{3}/2$).}
\end{figure}

With the numerical values $E = 1.25m \, , \, \kappa/\ell = \pm \sqrt{3}/2 \, \left(\beta = 0.5\right)$, the angular endpoints are at $\theta_\infty = 1.79\pi \left[\rm{rad}\right]$ for attraction ($\kappa < 0$) and at $\theta_\infty = 0.212\pi \left[\rm{rad}\right]$ for repulsion ($\kappa > 0$). Exemplary hodographs with these values, together with the corresponding spatial trajectories, are given in \Fref{fig:8} for attraction and \Fref{fig:7} for repulsion.

\vskip20pt

\section{Hodographs for exclusively relativistic cases} \label{sec: excrel}

The cases with $\ell \le |\kappa|$ don't have a nonrelativistic counterpart, because that would imply, as already pointed out, luminal or superluminal velocities in the Newtonian limit. The solutions for $\ell = |\kappa|$ and $\ell < |\kappa|$ are presented separately because of the different mathematical expressions, although the general features of the hodographs and the particle's spatial trajectories are similar in both cases.

\subsection{Hodographs for $\ell = |\kappa|$} \label{sec: elleq}

The hodograph solution in the light-like case is (with an arbitrary shift angle put to zero)
\begin{equation}\label{eq: ukeql}
\begin{eqalign}
{\vec u &= \frac{\epsilon E}{m} \theta \hat r -\epsilon \left( \frac{E}{2m} \theta^2 - \frac{E^2 - m^2}{2mE} \right) \hat \theta \\
 u^0 &= \frac{E^2 + m^2}{2mE} + \frac{E}{2m} \theta^2}
\end{eqalign}
\end{equation}
with $\epsilon = \rm{sign}\left( \kappa \right)$. It may be cast in the form of the general hodograph solution \eref{eq: genhod} as
\begin{equation}\label{eq: leqkhod}
 u^\mu \left( \theta |E,\ell \right) = \frac{m}{2E}v_o^\mu \left( \theta |E,\ell \right) + \frac{E}{2m} n^\mu \left( \theta |E,\ell \right)
\end{equation}
with the vector
\begin{equation}\label{eq: noleqk}
 n^\mu = \left( 1 + \theta^2, 2 \epsilon\theta \hat r + \epsilon \left( 1 - \theta^2 \right)\hat \theta \right) \, .
\end{equation}
Here both $v_o^\mu$ and $n^\mu$ are light-like, $v_o \cdot v_o = n \cdot n = 0$, and $n \cdot v_o = -2$.

\begin{figure}
  \centering
  \subfloat[Hodograph -- side view\label{fig:10a}]{\includegraphics[width=4.5cm]{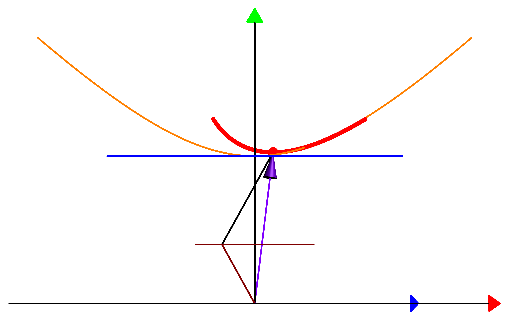}} \qquad
  \subfloat[Hodograph -- oblique view\label{fig:10b}]{\includegraphics[width=4.5cm]{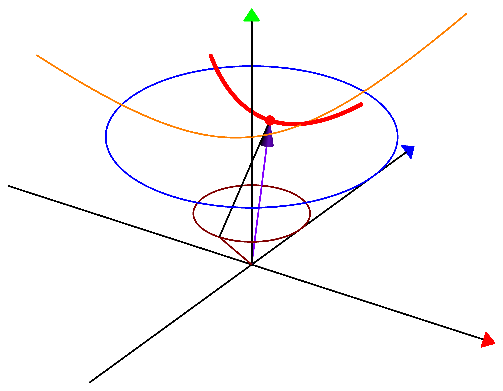}} \\
  \subfloat[Hodograph -- above view\label{fig:10c}]{\includegraphics[width=4.5cm]{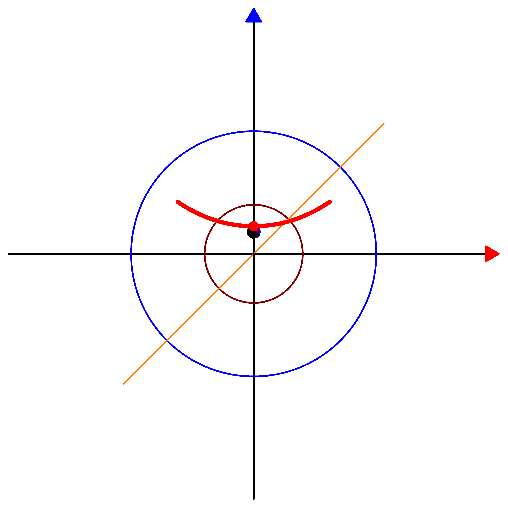}} \qquad
  \subfloat[Spatial orbit \label{fig:10d}]{\includegraphics[width=4.5cm]{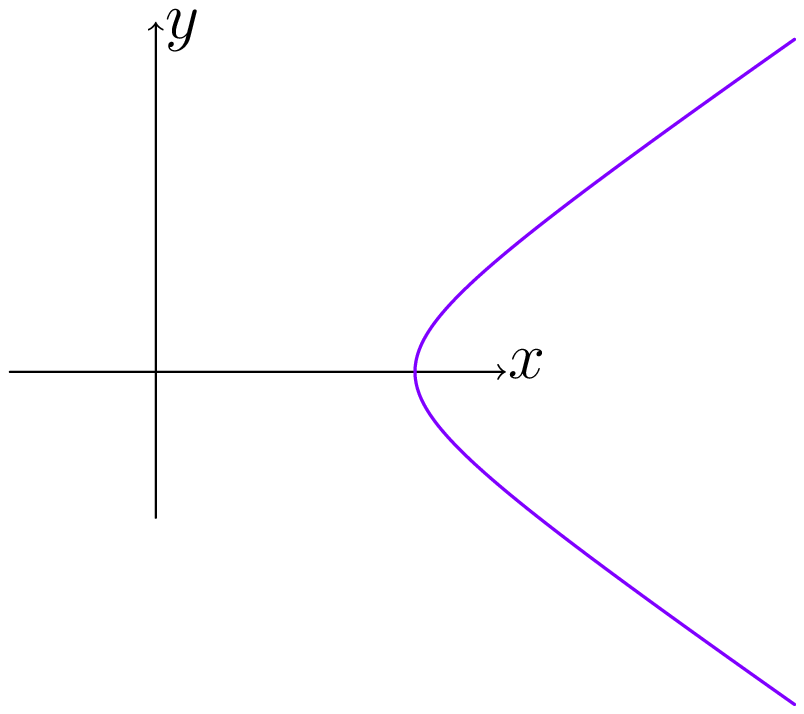}}
  \caption{\label{fig:10}Hodograph and spatial orbit for repulsion for $v_o^\mu$ light-like ($E = 1.25m, \ell = \kappa$).}
\end{figure}

For the angular ranges of the trajectory, determined by $u_\theta > 0$, we distinguish between the case of repulsion ($\kappa = \ell$ and $E > m$) for which
\begin{equation}\label{eq: thetarep}
 - \theta_{\infty} < \theta < \theta_{\infty}
\end{equation}
and two possible cases of attraction ($\kappa = - \ell$)
\begin{equation}\label{eq: thetatt}
 \cases{
 - \infty < \theta < \infty & when $\, E < m$ \\
\theta > \theta_{\infty} \quad {\rm or} \quad \theta < - \theta_{\infty} & when $\, E > m$ \\ }
\end{equation}
where (in both \eref{eq: thetarep} and \eref{eq: thetatt})
\begin{equation}\label{eq: thetainfleqk}
 \theta_{\infty} = \frac{\sqrt {E^2 - m^2}}{E} \, .
\end{equation}
\begin{figure}
  \centering
  \subfloat[Hodograph -- side view \label{fig:12a}]{\includegraphics[width=4.5cm]{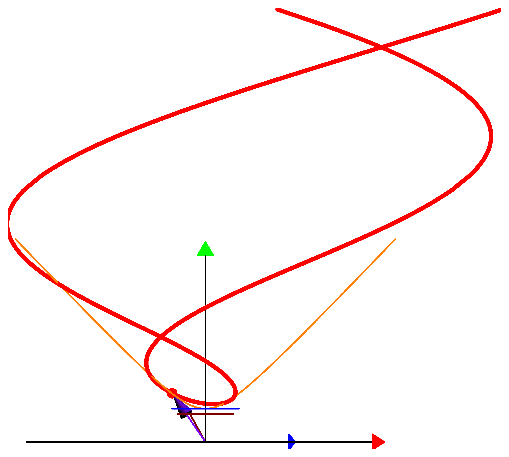}} \qquad
  \subfloat[Hodograph -- oblique view \label{fig:12b}]{\includegraphics[width=4.5cm]{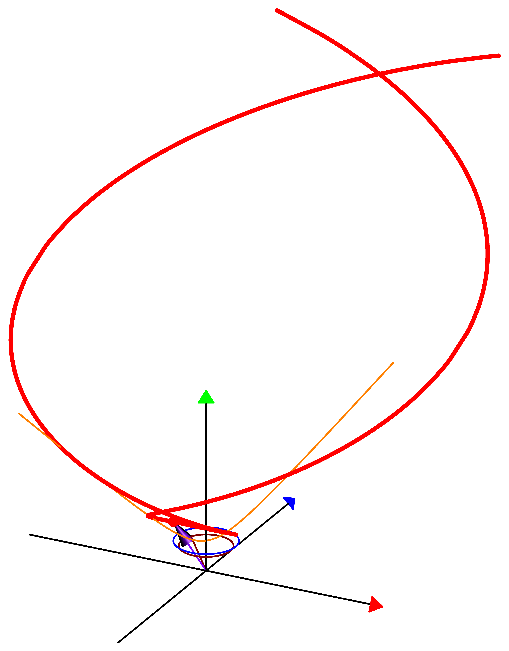}} \\
  \subfloat[Hodograph -- above view \label{fig:12c}]{\includegraphics[width=4.5cm]{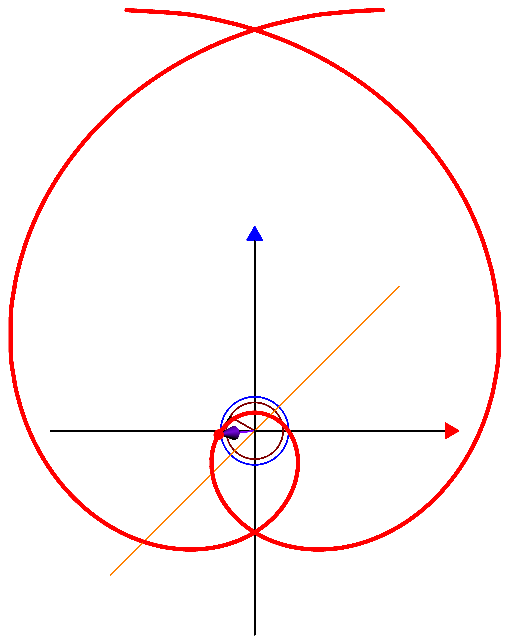}} \qquad
  \subfloat[Spatial orbit \label{fig:12d}]{\includegraphics[width=4.5cm]{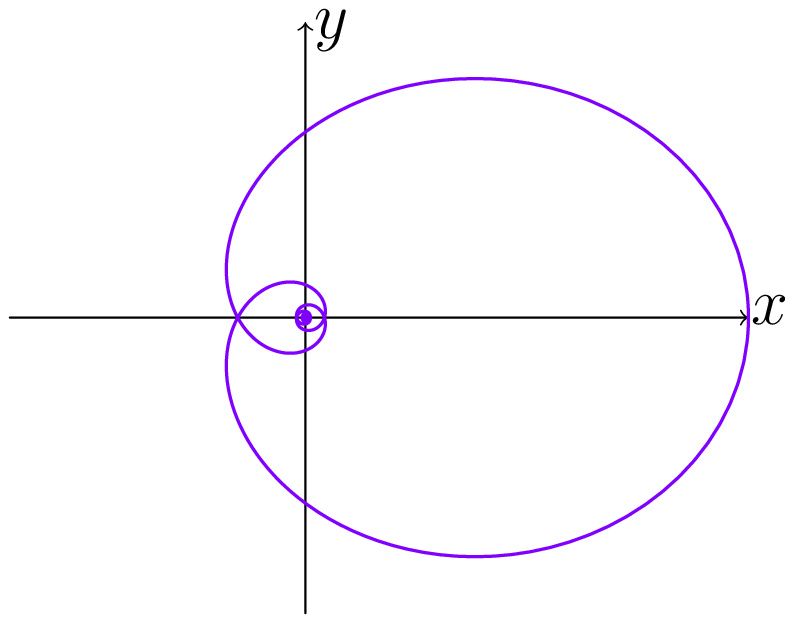}}
  \caption{\label{fig:12} Hodograph and spatial orbit for 1st attraction (unstable bound-like) case for $v_o^\mu$ light-like ($E = 0.6m, \ell = -\kappa$).}
\end{figure}

In the case of repulsion ($\kappa = \ell$) the particle comes from infinity ($r \rightarrow \infty$ , $u_\theta = 0$) to a minimal distance at $\theta = 0$
 \begin{equation} \label{eq: rminkeql}
 r_{\rm min} = \frac{\ell}{m u_{\theta{\rm max}}} = \frac{2\ell E}{E^2 - m^2}
 \end{equation}
and then back again to infinity. The hodograph is a finite segment between $u^\mu_{-\infty}$ and $u^\mu_{\infty}$, where
\begin{equation}\label{eq: replimits}
 u^\mu_{\pm\infty} = u^\mu\left(\pm\theta_{\infty}\right) = \left( \frac{E}{m}, \pm\frac{\sqrt{E^2 - m^2}}{m} \hat r\left(\pm\theta_{\infty}\right) \right)
\end{equation}
For the particular choice $E = 1.25m$ the endpoints are at $\theta_\infty = 0.6\left[\rm{rad}\right]$. The hodograph for these values, together with the corresponding spatial trajectory, is given in \Fref{fig:10}.

\begin{figure}
  \centering
  \subfloat[Hodograph -- side view \label{fig:14a}]{\includegraphics[width=4.5cm]{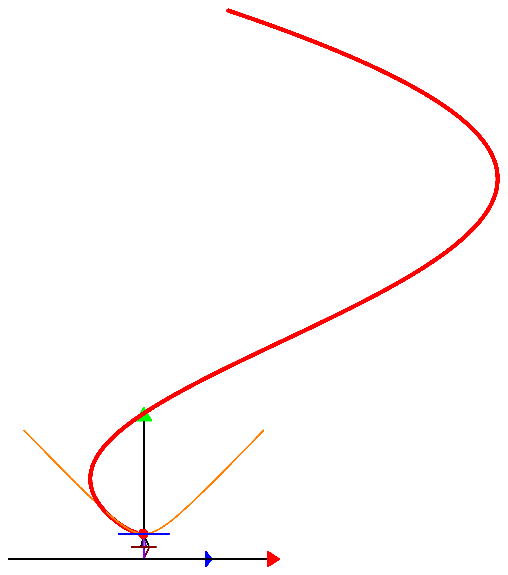}} \qquad
  \subfloat[Hodograph -- oblique view \label{fig:14b}]{\includegraphics[width=4.5cm]{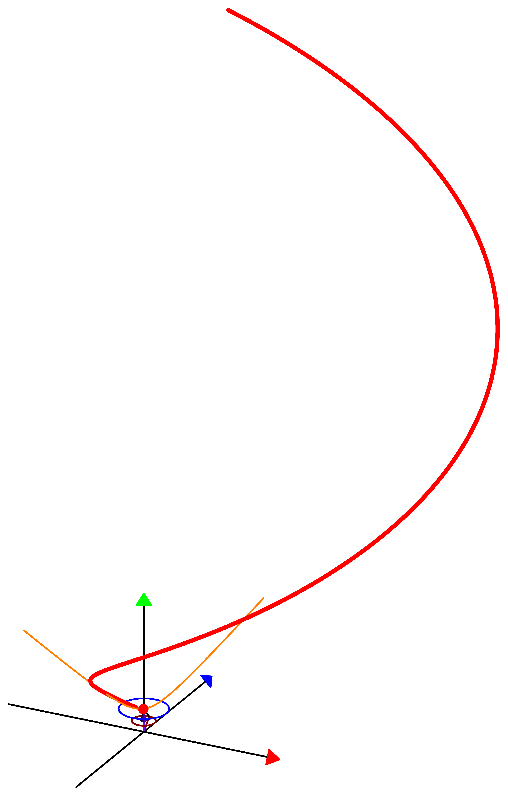}} \\
  \subfloat[Hodograph -- above view \label{fig:14c}]{\includegraphics[width=4.5cm]{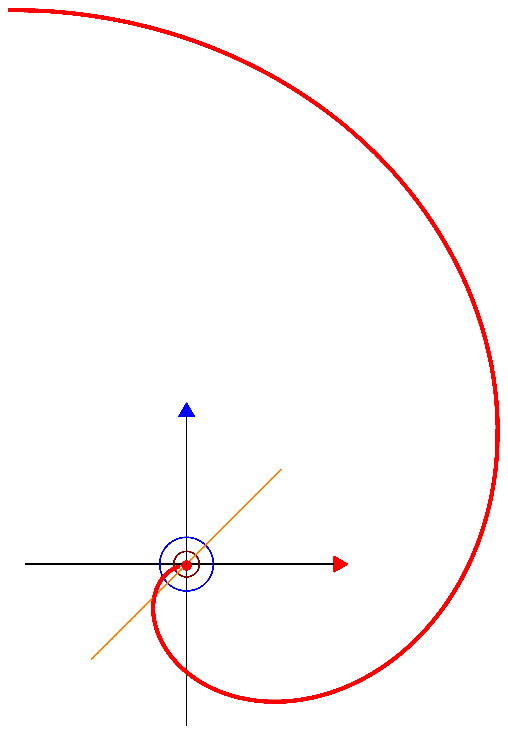}} \qquad
  \subfloat[Spatial orbit \label{fig:14d}]{\includegraphics[width=4.5cm]{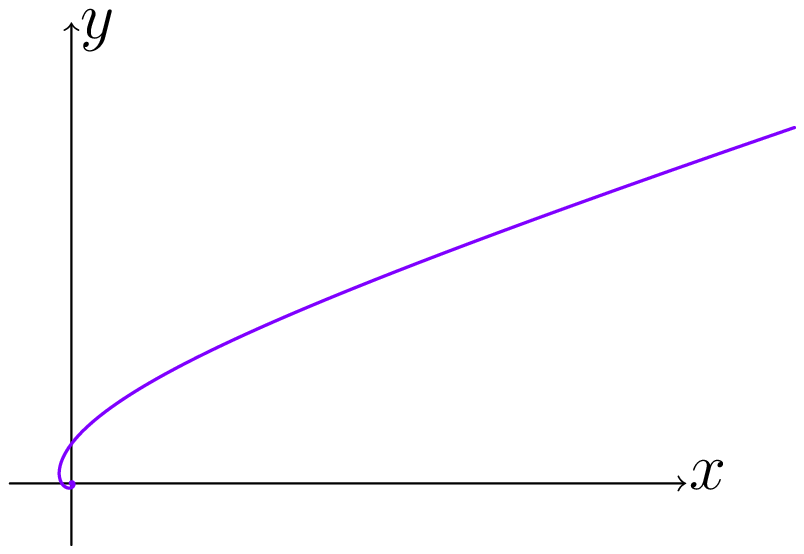}}
  \caption{\label{fig:14} Hodograph and spatial orbit for 2nd attraction (collapse) case for $v_o^\mu$ light-like ($E = 1.05m, \ell = -\kappa$).}
\end{figure}

The first attraction case ($\kappa = -\ell \, , \, E < m$) resembles an unstable bound state : $\theta$ may get any value without bounds ($- \infty < \theta < \infty$), but the particle is spatially bound within the maximum distance from the centre
\begin{equation}\label{eq: maxrleqk}
 r_{\max} = \frac{\ell}{m u_{\theta \min}} = \frac{2E\ell}{m^2 - E^2}
\end{equation}
The particle may start spatially at some distance from the centre of force, then move in a spiral trajectory towards the centre. At the same time the hodograph starts close to the centre of velocity space and spirals away to infinity (velocity space infinity). In the limit $\theta \to \infty$ then also $\vec u \to \infty$ ($\left|\vec v\right| \to c$), and the particle collapses into the centre of force. The hodograph with the choice $E = 0.6m$, together with the corresponding spatial trajectory, is given in \Fref{fig:12}.

In the second attraction case ($\kappa = -\ell \, , \, E > m$) a situation similar to gravitational singularity is encountered : Either (i) $\theta_\infty < \theta < \infty$, then the hodograph ia a semi-infinite segment, the particle arrives from spatial infinity ($u_\theta = 0$) and collapses into the centre of force for $\theta \to \infty$ ($u_\theta \to \infty$, $r \to 0$), or (ii) $ -\infty < \theta < - \theta_\infty$, again a semi-infinite hodograph, and the particle bursts out of the centre of force corresponding to $\theta \to -\infty$ ($u_\theta \to \infty$, $r \to 0$) and escapes to infinity ($u_\theta \to 0$) (although the latter possibility is bizarre and unlikely, it is mentioned because of its theoretical possibility). The hodograph with the choice $E = 1.05m$, for which $\theta_\infty = 0.01\left[\rm{rad}\right]$, together with the corresponding spatial trajectory, is given in \Fref{fig:14}.

\subsection{Hodographs for $\ell < |\kappa|$} \label{sec: ellsmall}

\begin{figure}
  \centering
  \subfloat[Hodograph -- side view \label{fig:9a}]{\includegraphics[width=4.5cm]{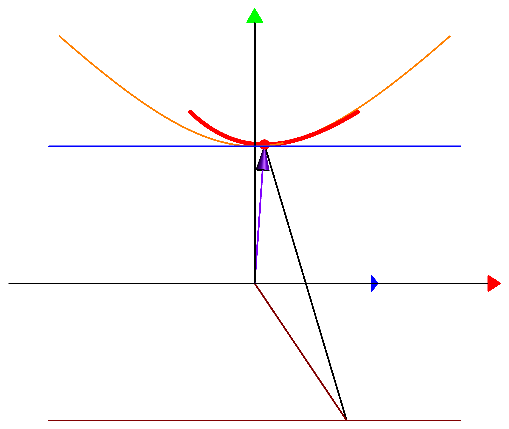}} \qquad
  \subfloat[Hodograph -- oblique view \label{fig:9b}]{\includegraphics[width=4.5cm]{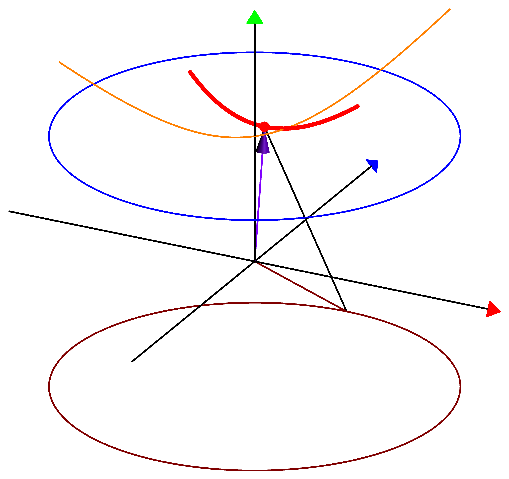}} \\
  \subfloat[Hodograph -- above view \label{fig:9c}]{\includegraphics[width=4.5cm]{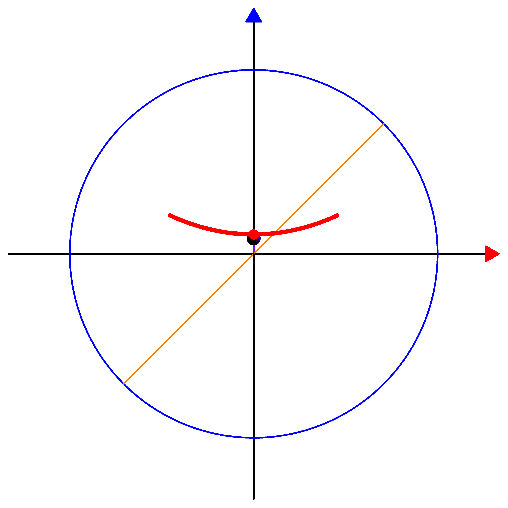}} \qquad
  \subfloat[Spatial orbit \label{fig:9d}]{\includegraphics[width=4.5cm]{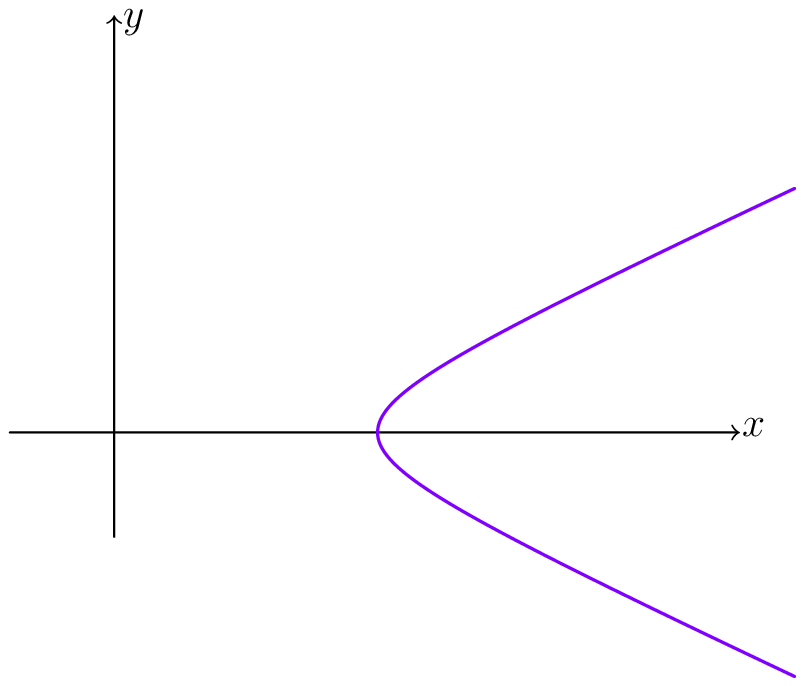}}
  \caption{\label{fig:9} Hodograph and spatial orbit for repulsion for $v_o^\mu$ space-like ($E = 1.25m, \kappa/\ell = 1.5$).}
\end{figure}
\begin{figure}
  \centering
  \subfloat[Hodograph -- side view \label{fig:11a}]{\includegraphics[width=4.5cm]{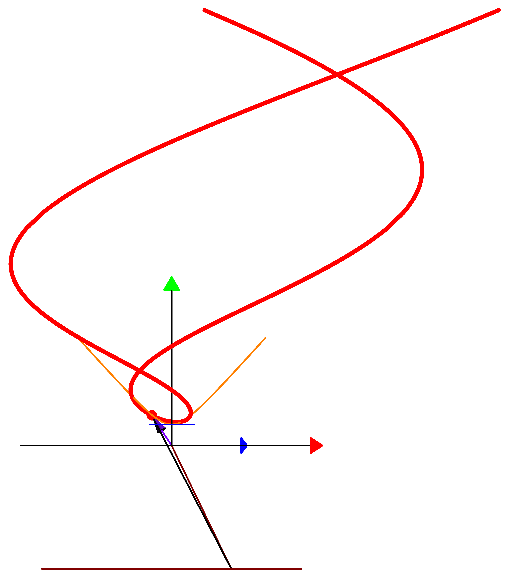}} \qquad
  \subfloat[Hodograph -- oblique view \label{fig:11b}]{\includegraphics[width=4.5cm]{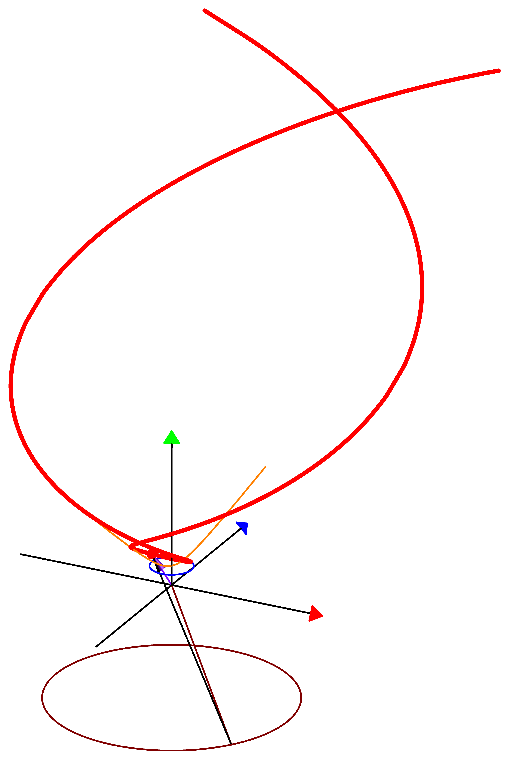}} \\
  \subfloat[Hodograph -- above view \label{fig:11c}]{\includegraphics[width=4.5cm]{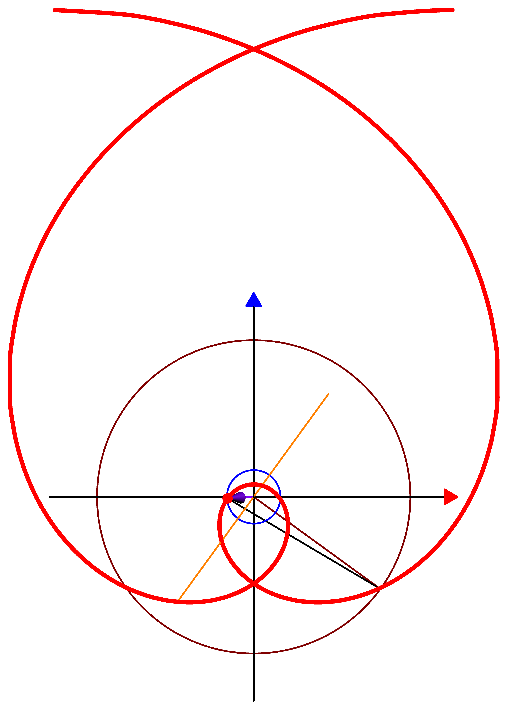}} \qquad
  \subfloat[Spatial orbit \label{fig:11d}]{\includegraphics[width=4.5cm]{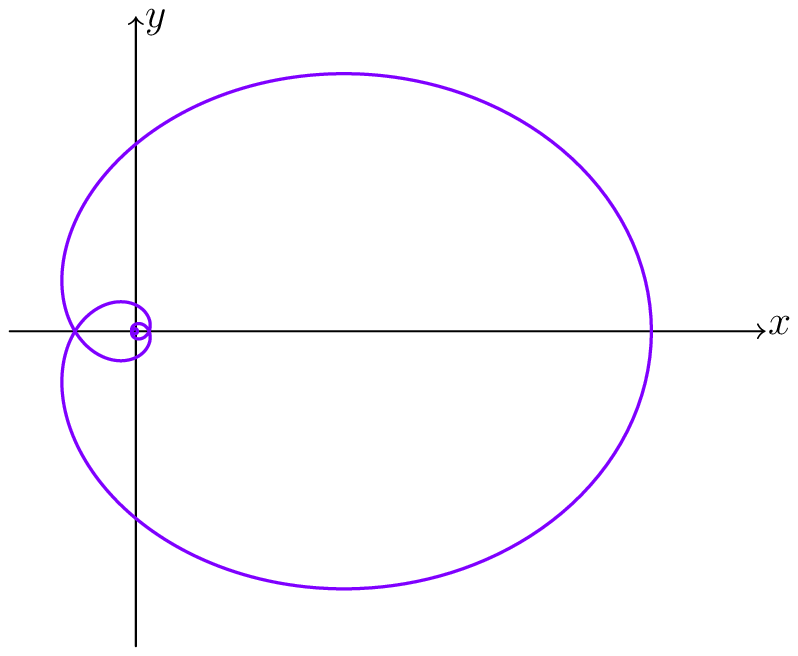}}
  \caption{\label{fig:11} Hodograph and spatial orbit for 1st attraction (unstable bound-like) case for $v_o^\mu$ space-like ($E = 0.6m, \kappa/\ell = -1.05$).}
\end{figure}

\begin{figure}
  \centering
  \subfloat[Hodograph -- side view \label{fig:13a}]{\includegraphics[width=4.5cm]{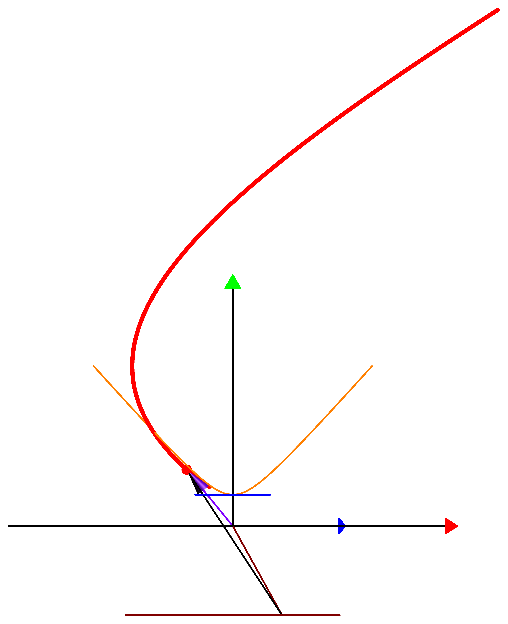}} \qquad
  \subfloat[Hodograph -- oblique view \label{fig:13b}]{\includegraphics[width=4.5cm]{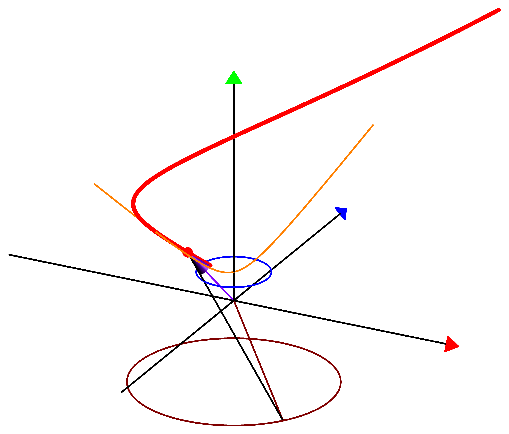}} \\
  \subfloat[Hodograph -- above view \label{fig:13c}]{\includegraphics[width=4.5cm]{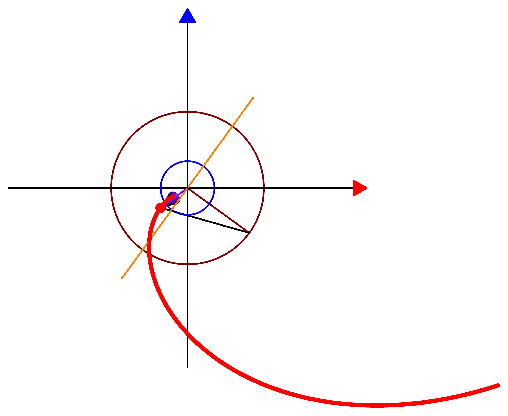}} \qquad
  \subfloat[Spatial orbit \label{fig:13d}]{\includegraphics[width=4.5cm]{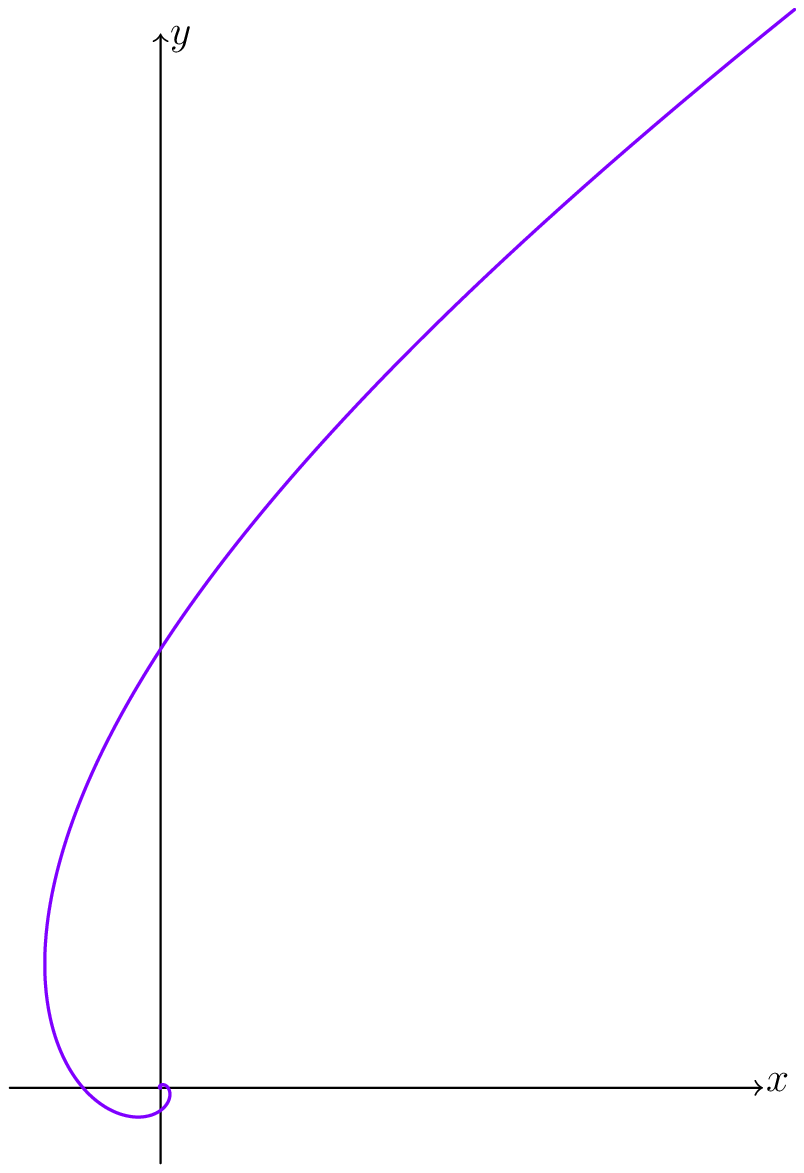}}
  \caption{\label{fig:13} Hodograph and spatial orbit for 2nd attraction (collapse) case for $v_o^\mu$ space-like ($E = 1.25m, \kappa/\ell = -1.2$).}
\end{figure}

The hodograph solution in the space-like case (with an arbitrary shift angle put to zero) is
\begin{equation}\label{eq: uklarge}
\begin{eqalign}{
 \vec u &= - \epsilon A_o \sinh \left(\bar\beta \theta \right) \hat r + \left[ \frac{\kappa E}{m\ell \bar\beta^2} - \frac{\epsilon A_o}{\bar\beta} \cosh \left(\bar\beta \theta \right) \right] \hat \theta \\
 u^0 &= - \frac{E}{m \bar\beta^2} + \frac{\left| \kappa \right| A_o}{\bar\beta \ell} \cosh \left(\bar\beta \theta \right)}
\end{eqalign}
\end{equation}
where $\epsilon = \rm{sign} \left( \kappa \right)$ and with the notations
\begin{equation}\label{eq: Aobetabar}
 \bar\beta = \sqrt {\frac{\kappa^2}{\ell^2} - 1} \quad , \quad A_o = \sqrt {\frac{E^2}{m^2 \bar\beta^2} + 1} = \frac{\Lambda\left(E,\ell\right)}{m \bar\beta} \, .
 \end{equation}
As with the previous cases, it may be cast into the general hodograph form \eref{eq: genhod} as
\begin{equation}\label{eq: klargehod}
 u^\mu \left(\theta | E,\ell \right) = - \frac{E}{m \bar\beta^2} v_o^\mu + A_o n^\mu
\end{equation}
with
\begin{equation}\label{eq: wnklarge}
 n^\mu \left(\theta | \ell \right) = \left( \frac{\left| \kappa \right|}{\bar\beta \ell} \cosh \left( \bar\beta \theta \right), \epsilon \sinh \left( \bar\beta \theta \right) \hat r - \frac{\epsilon}{\bar\beta} \cosh \left( \bar\beta \theta \right) \hat \theta \right)
\end{equation}
The r\^{o}les of $v_o^\mu$ and $n^\mu$ are now interchanged, since $v_o^\mu$ is space-like in the present case and $n^\mu$ is a time-like unit 4-vector. The negative coefficient of $v_o^\mu$ turns it up-side down, pointing downwards rather than upwards. This behaviour is clear in Figures \ref{fig:9}, \ref{fig:11} \& \ref{fig:13}. Although this case doesn't have a Newtonian counter-part, the composition in \eref{eq: klargehod} suggests, in analogy with \eref{eq: usol2}, to regard the vector $A_o n^\mu$ as the Hamilton vector in the present cases.

The angular ranges of the possible trajectories are, in principle, the same as in the former case, given by \eref{eq: thetarep} for repulsion and \eref{eq: thetatt} for attraction, this time with
\begin{equation}\label{eq: thetainflargek}
 \theta_{\infty} = \frac{1}{\bar\beta} \cosh^{- 1} \left( \frac{\kappa E}{\ell \Lambda}\right) \, .
\end{equation}
In the case of repulsion ($\kappa > \ell$ and $E > m$), with the angular range $- \theta_{\infty} < \theta < \theta_{\infty}$, the particle comes from spatial infinity to closest approach at $\theta = 0$, with the minimal distance given in this case by
\begin{equation}\label{eq: minrlargek}
 r_{\min} = \frac{\ell}{m u_{\theta\max}} = \frac{\ell\Lambda + \kappa E}{E^2 - m^2}
\end{equation}
and then back again to infinity. The hodograph is a finite segment between $u^\mu_{-\infty}$ and $u^\mu_{\infty}$, with $u^\mu_{\pm\infty}$ given again by \eref{eq: replimits}. For the particular choice $E = 1.25m$ the endpoints are, again, at $\theta_\infty = 0.6\left[\rm{rad}\right]$. An hodograph for these values and $\kappa = 1.5\ell$, together with the corresponding spatial trajectory, is given in \Fref{fig:9}.

The first case of attraction in \eref{eq: thetatt} ($\kappa < - \ell \, , \, E < m$), with the angular range $- \infty < \theta < \infty$, again resembles an unstable bound state with the maximum distance from the centre of force
\begin{equation}\label{eq: maxrlargek}
 r_{\max} = \frac{\ell}{m u_{\theta \min}} = \frac{\ell\Lambda + \left|\kappa\right|E}{m^2 - E^2}
\end{equation}
The particle starts spatially at some distance from the centre, then moves in a spiral trajectory towards the centre, eventually collapsing into it. The hodograph starts close to the centre of velocity space and spirals away to (velocity space) infinity. These hodographs are illustrated, with the choice $E = 0.6m$ and $\kappa = -1.05\ell$, together with the corresponding spatial trajectory, in \Fref{fig:11}.

In the second attraction case ($\kappa < - \ell \, , \, E > m$) then, again, as with the light-like case : Either (i) $\theta_\infty < \theta < \infty$, then the hodograph is a semi-infinite segment, the particle arrives from spatial infinity ($u_\theta = 0$) and collapses into the centre of force for $\theta \to \infty$ ($u_\theta \to \infty$, $r \to 0$), or (ii) $ -\infty < \theta < - \theta_\infty$, again a semi-infinite hodograph where the particle bursts out of the centre of force corresponding to $\theta \to -\infty$ ($u_\theta \to \infty$, $r \to 0$) and escapes to infinity ($u_\theta \to 0$). These hodographs are illustrated with the choice $E = 1.25m$, for which $\theta_\infty = 0.6\left[\rm{rad}\right]$, and $\kappa = -1.2\ell$, together with the corresponding spatial trajectory, in \Fref{fig:13}.

\vskip20pt

\section{Concluding remarks} \label{sec: Conrem}

The purpose of the talk was mainly to present the application of the generally non-familiar {\it hodograph method} -- studying the dynamics of a system in velocity space -- as it applies to a charged relativistic particle in Coulomb field. The hodograph method has the merit that when applied to systems with $1/r$ potentials, Newtonian and relativistic alike, the velocity equations are simply linear, providing a very straight-forward and elegant means to analyze the dynamics of the system.

The Newtonian Kepler/Coulomb (KC) hodograph is always a circle, in general displaced from the origin of velocity space, with a constant displacement that depends on the energy state and determines the shape of the (conic section) spatial trajectory. For relativistic Coulomb systems this structure is preserved, in principle -- a basic circle associated with the minimum energy state, uniformly displaced depending on the actual energy state. The unique relativistic feature is that the displacement is itself rotating, so the whole phenomenon is of rotation superimposed on rotation -- precession on the velocity hyperboloid -- thus the title with ``back to epicycles".

Results of the analysis were presented in the form of an excursion in the garden of the many possible manifestations of the system, mostly in an illustrative way. Explicit distinction was made between the cases which have a non-relativistic limit and others which are exclusively relativistic, demonstrating the particular characteristics of each.

This analysis, whose details will be published elsewhere, is sought to assist attempts to advance the solution of the relativistic EM 2-body problem : Unlike the Newtonian case, the relativistic (non-quantum) EM 2-body problem doesn't have, despite all the years, a satisfactory solution. Some simple particular solutions have been found, mainly in the 60's and the 70's \cite{Stephas1978}, but since then no real advancement has been marked and a general solution is missing.

However, as is well known, the $1/r$ interaction allows Newtonian KC systems to enjoy a very special and simple form of the Laplace-Runge-Lenz (LRL) symmetry, which is also directly associated with the spatial trajectories being conic sections \cite{GoldsteinLRL,Goldstein.etal2000,KhachidzeKhelash08}. The combination of rotational symmetry with the extra LRL symmetry provides the full solution for these systems in configuration space. The application of the hodograph method to Newtonian KC systems, providing the full solution in velocity space, may be demonstrated to be equivalent with the application of the LRL symmetry in ordinary space. In particular, the vector displacing the basic hodograph circle in velocity space, the Hamilton vector, is a constant vector directly related to the constant LRL vector in the same system \cite{Munoz2003}.

The hodograph method therefore displays what may be termed {\it Hamilton's symmetry}, complementary to the LRL symmetry. These symmetries are mathematically equivalent, but physically they behave differently : the LRL symmetry transforms between states of same energy but different internal angular momentum, while the Hamilton symmetry transforms between states of same angular momentum while changing the energy.

The success of using the extra symmetries with Newtonian KC systems gives rise to the hope that by studying the corresponding extra symmetries in relativistic EM systems a general solution may be advanced. It is in this context that the present work has been performed.

Newtonian KC 2-body systems may be reduced to 1-body systems by transition to the centre-of-mass (CM) reference frame. Such a simple and direct procedure is impossible, in general, for relativistic 2-body systems because the interaction is not instantaneous. It is therefore convenient to start with Coulomb systems, which may be regarded as the limit of EM 2-body systems when one of the charges is much heavier than the other. The LRL symmetry in relativistic Coulomb systems was already studied to some extent in recent years \cite{Stahlhofen2005a,Yoshida1988,LRLgen}. Here we considered the other face of these internal symmetries -- the relativistic Hamilton symmetry -- demonstrating its usefulness for Coulomb systems.

In futuristic view, an answer to the question {\it How may non-instantaneous interactions be handled on velocity space, especially in view of the fact that time doesn't appear in the structure of the velocity space ?} may provide a key to the whole issue of the 2-body EM problem. It is hoped to come back to such questions soon.

\vskip20pt

\rule{10cm}{1pt}


\end{document}